\documentclass[letterpaper,twocolumn,prb]{revtex4}

\usepackage{url}
\usepackage[dvips]{graphicx}

\newcommand{\Journal}[4]{\textit{#1} \textbf{#2}, #3 (#4)}

\begin{document}

\title{What can we learn about neutron stars from gravity-wave
observations?\footnote{Talk presented at the 25th J.\ Hopkins Workshop on Current Problems in Particle Theory, \emph{2001: A Relativistic Spacetime Odyssey}, Florence, Sep.\ 3--5, 2001.}}

\author{Michele Vallisneri}

\affiliation{Theor.\ Astrophys.\ 130-33, California Institute of
Technology, Pasadena, CA 91125}

\date{Jan 11, 2002}

\begin{abstract}
In the next few years,
the first detections of gravity-wave signals using Earth-based interferometric detectors will begin to provide precious new information about the structure and dynamics of compact bodies such as neutron stars.  The intrinsic weakness of gravity-wave signals requires a proactive approach to modeling the prospective sources and anticipating the shape of the signals that we seek to detect.  Full-blown 3-D numerical simulations of the sources are playing and will play an important role in planning the gravity-wave data-analysis effort.  I review some recent analytical and numerical work on neutron stars as sources of gravity waves.
\end{abstract}

\maketitle

\section{Introduction}
In the course of the next decade, the inception of gravity-wave (GW)
astronomy will open an exciting new window on the physics of compact,
strongly gravitating objects such as neutron stars (NSs) and black holes (BHs), providing
information complementary to that available from electromagnetic and neutrino observations, and plausibly producing important insights
into unsolved questions such as the equation of state (EOS) of matter at
nuclear densities, and the mechanisms behind gamma-ray
bursts. Detailed reviews of GW sources, of the expected event rates,
and of the physics that these sources could teach us are available
elsewhere,\cite{Ferrari,Hughes,Kip} but I shall list briefly the most
promising astrophysical systems from which we could learn about NSs
using Earth-based GW interferometers such as LIGO and VIRGO.
\begin{enumerate}
\item \emph{NS--NS and NS--BH binaries in the last few minutes of
their inspirals.}  For a long time, these inspiraling systems have been
the prototype for the category of the short-lived \emph{chirp} signals detectable using Earth-based interferometers. The reason, of course,
is that NS--NS binaries have actually been observed in our
galaxy,\cite{Thorsett} but also that the part of the inspiral
accessible to the interferometers (with GW frequencies between 40 and
1000 Hz) sits well before the final merger of the binary, so it is
described very accurately by the well-developed post--Newtonian
equations for point masses.\cite{Blanchet} The successful
observation of GWs from these events will teach us about the masses,
spins and locations of NSs, but not about their internal structure.

On the contrary, the detection of GW from the endpoint of NS--BH inspirals should produce detailed information about NS structure and EOS. For a wide range of binary parameters, the NS will be torn apart by the tidal field of the BH well before the final plunge into the hole, and the tidal-disruption waves will be well inside the frequency range of good interferometer sensitivity.\cite{Vallisneri} NS--BH binaries have also been proposed as engines for gamma-ray bursts\cite{Janka} and as suitable environments for the production of heavy nuclei in $r$-processes.\cite{Lee} I will discuss these systems more extensively in \mbox{Sec.\ \ref{sec:tidal}} of this paper.
\item \emph{Rapidly spinning, deformed NSs.}  This class includes the
known and unknown radio pulsars (when their gravitational
ellipticity is high enough to provide strong GWs), and the systems
known as low-mass X-ray binaries (LMXB), where the NS is accreting
matter and angular momentum from a companion, but instead of
increasing its rotation, it is locked into spin periods of about 3 ms;
it is conjectured that the angular momentum being accreted is
lost to the emission of GW.\cite{Bildsten}

To detect rapidly spinning NSs, it
will be necessary to integrate the GW signal for times up to several
months, so the Doppler frequency modulation caused by the movement of
the Earth around the sun will make it much harder to detect previously
unknown sources.\cite{Brady} At the same time, the shape of this
modulation will make it possible to obtain the position of the source
in the sky,\cite{Jaranowski} and to match the GW source with one of
the objects known from electromagnetic observations.

If any GWs are detected, their features would be very informative, in particular when examined in correlation with electromagnetic signals from the same source. For instance, the ratio of the GW frequency to the NS angular frequency could identify the nature of the inhomogeneities that give rise to the GW emission, and the evolution of the GW amplitude and frequency could provide interesting data about NS physics such as crust structure and dynamics, crust--core interactions, magnetic fields, viscosity, superfluidity, and more.\cite{Kip}
\item \emph{Proto-neutron stars.}  Finally, NSs could be
observed as the rapidly spinning, strongly asymmetric remnants of
stellar-core collapse, or as the proto--NSs produced by the
accretion-induced collapse of white dwarves. Proto--NSs that
spin very fast can hang up centrifugally at a stage where their radius
is still large compared to that of the final NS. Such a
configuration might be unstable to a bar mode, giving rise to an
elongated object that would emit very strong GWs.\cite{bars}
The newborn NSs might also develop a GW-induced instability in
their $r$-modes.\cite{Lindblom,Kokkotas} I will discuss this
possibility more extensively in Sec.\ \ref{sec:rmode}.

For all these systems, GWs would provide information complementary to that made available by neutrino observations, focusing on the density structure and asymmetry of the collapsing core rather than on its thermal structure.
\end{enumerate}

\noindent While we wait eagerly for the first detections, numerical
simulations of GW sources are beginning to provide precious insight into
the effect of NS structure and dynamics on the GW
signals. The purpose of this paper is to explore how 
simulations are helping and directing the theory and practice of GW
data analysis. To do so, I will single out two promising GW sources
(NS tidal disruption in NS--BH binaries, in Sec.\ \ref{sec:tidal}; NS $r$-modes, in Sec.\ \ref{sec:rmode}) that are currently being
attacked with numerical simulations, and on which I have first-hand
experience.

\section{Neutron-star tidal disruption as a probe into the equation of
state of dense nuclear matter}
\label{sec:tidal}

With event rates between $10^{-7}$ and $10^{-4}$ per year in our
galaxy, NS--BH mergers are one of the standard GW sources for
second-generation interferometers (LIGO-II should be able to detect
these systems out to 650 Mpc, yielding 1--1500 observations per
year\cite{Kip}).  The waveforms generated by these events will contain
two kinds of information. The early part of the inspiral (during which
the NS and BH are still relatively distant, and the dynamics can be
described accurately by post--Newtonian equations of motion in the
point-mass approximation) will tell us about the masses and the spins
of the NS and BH.  The late part of the inspiral, depending on the
binary parameters, can see the BH tidal field become so strong that it
disrupts the NS on a dynamical timescale; physical intuition then
suggests that the details of the disruption process, as encoded in
GWs, should carry useful information about the internal structure of
the NS, and in particular about its EOS.\cite{theidea}

\subsection{A simple analytical model for NS tidal disruption}

The prospects for extracting this information from the GWs that could be measured from a realistic event have been evaluated by the present author,\cite{Vallisneri} using a crude model that however accounts for most of the relevant physics.  The simplest possible representation of a NS inspiraling into a BH is a quasi-equilibrium sequence of \emph{relativistic Roche--Riemann ellipsoids}. These ellipsoids are equilibrium configurations of a self-gravitating, polytropic, Newtonian fluid, moving on circular, equatorial geodesics in the Kerr spacetime, and subject to the BH relativistic tidal field.\cite{Chandra} For these configurations, once the binary parameters $m$, $M$, $a$ and $r$ (respectively, the NS and BH masses, the BH spin, and the binary separation) are fixed, there is only one free parameter, corresponding to the NS radius $R$. So we will take $R$ as a representative of the uncertainty in the NS EOS.

\begin{figure}
\begin{center}
\includegraphics[width=3.3in]{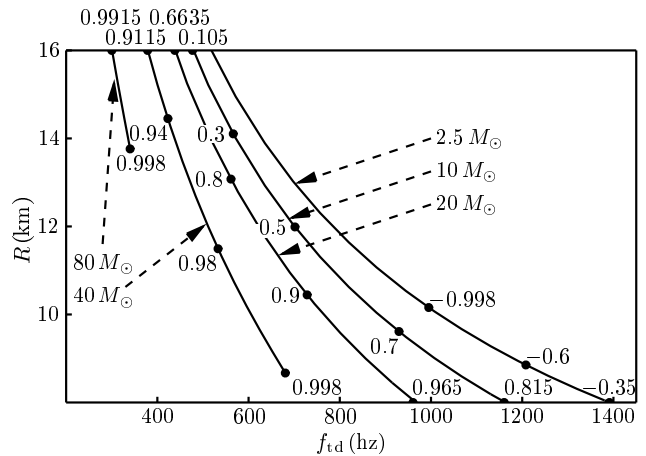}
\caption{GW frequency at tidal disruption, $f_\mathrm{td}(m,R,M,a)$, as
a function of NS radius $R$. Here we have set the NS mass to $1.4 \,
M_\odot$, we have indicated the BH mass on the curves, and we have
omitted the dependence on the BH spin $a$, which gives negligible
corrections.  The BH spin parameter $a$ is nevertheless important to
decide whether the NS will actually disrupt before it begins its rapid
plunge into the BH. Estimating the onset of the plunge as the naive $6
M$ ISCO, we find that to have disruption we need $a/M$ to assume at least the value indicated by the big dots in this figure. Negative $a/M$
denotes counterrotating orbits.\label{fig:disrupt}}
\end{center}
\end{figure}

As it inspirals toward the BH, a NS with parameters $m$ and $R$
would be represented by the appropriate Roche--Riemann ellipsoid at
each separation $r$, until we reach a \emph{critical} $r_\mathrm{cr}$,
beyond which no more equilibrium configurations
exist.  We identify this end of the
equilibrium sequence with the onset of dynamical tidal disruption, and
from $r_\mathrm{cr}$ we obtain the GW frequency at tidal disruption,
$f_\mathrm{td} = f_\mathrm{td}(m,R,M,a)$.  This function is shown in
Fig.\ \ref{fig:disrupt}, for the standard NS mass $m=1.4 \, M_\odot$,
and for a variety of BH masses.  Figure \ref{fig:disrupt} should be
read as follows: given a NS--BH signal, choose the curve corresponding to the BH mass (as estimated from the frequency evolution of the early inspiral signal);
on the horizontal axis, locate the frequency at which tidal disruption
begins (as estimated from the late part of the GW signal); then read off
the estimated NS radius on the vertical axis.

Figure \ref{fig:disrupt} suggests that the GW frequency at tidal
disruption depends strongly on the NS radius, and that the disruption
waveforms lie in the band of good interferometer sensitivity for the
advanced interferometers such as LIGO-II.  It follows that, in
principle, we could use the waveforms from a NS tidal-disruption event
to measure both the NS mass and the NS radius. However, we still need to know just how well we could measure them.

\subsection{The matched-filtering paradigm for gravity-wave data analysis}

To answer this question, we need to invoke the theory of
\emph{matched-filtering parameter estimation}.\cite{parest} The
general idea is that GWs will be detected by correlating the measured
signal $|s\rangle$ to a bank of theoretical \emph{templates}
$\{|h_i\rangle\}$, which represent our best approximation of the
realistic GW signal, as it could happen for a variety of binary
parameters.  If the \emph{match} $\langle s| h_i\rangle$ (the
correlation between $|h_i\rangle$ and $|s\rangle$) is much higher than
the match $\langle h_i|n\rangle$ that the template would give, on the
average, with noise alone, then we claim that we have a detection.  To
know how well we can estimate $R$, we ask how probable it is that a
particular realization of detector noise would lead us to mistake the
template $|h_R\rangle$ with the nearby template $|h_{R+\Delta
R}\rangle$: the answer is given in terms of the match $\langle
h_{R}|h_{R+\Delta R}\rangle$.

To compute this match we construct a bank of signal templates
that differ only\footnote{In the realistic case, the templates depend
on \emph{all} the binary parameters.  The estimation problem then
becomes involved, because there can be correlations in the ways that
different parameters modify the templates.  In our simple model, we
assume that all parameters except $f_\mathrm{td}$ are already known
well from of the early part of the inspiral signal, so all that is
left to do is to find $f_\mathrm{td}$.} in the GW frequency at the
onset of tidal disruption.  Because we are not interested in modeling
accurately the relativistic dynamics of the binary, but only the
effects of tidal disruption, we generate our waveforms from simple
quadrupole-governed Newtonian inspirals,\cite{mtw} cutting off the signal, more
or less abruptly,\footnote{Bildsten and Cutler\cite{Cutler} estimate
that complete disruption would take place in $\sim$ 1--3 orbital
periods, while the disrupted NS would spread into a ring in $\sim$
1--2 periods, significantly reducing the GW amplitude.} when the
instantaneous GW frequency reaches $f_\mathrm{td}$. Computing the
match between nearby templates, we estimate the granularity to
which $f_\mathrm{td}$ can be measured for a given signal strength
(inversely proportional to distance); we can then use our equation for
$f(R)$ to propagate the errors to the NS radius.  The final result of
this exercise is that, employing advanced Earth-based interferometers
such as LIGO-II\cite{Kip}, we should be able to measure the NS radius
to 15\%, for tidal-disruption events at distances that yield about one
event per year.\cite{Vallisneri} This estimated error should be compared to the error (about a factor of two) in the measurements of the NS
\emph{thermal radius}.

\subsection{Inverting the mass--radius relation}

What can we do once we get a few points on the $m(R)$ curve?  We can
try to solve for the EOS of dense nuclear matter in NSs. The
relativistic model of nonrotating NSs may be considered as a mapping
from the NS EOS, through the Oppenheimer--Volkoff (OV) equations,
\begin{equation}
\frac{dm}{dr} = 4 \pi r^2 \rho, \quad
\frac{d\rho}{dr} = -(\rho + p)\frac{m + 4 \pi r^3 p}{r(r-2m)},
\end{equation}
to macroscopic NS quantities such as mass and radius. Let us work
through this mapping in the case of a polytropic EOS, $p =
p(\rho)$. First, we set the central density $\rho_c$; then, we solve
the OV equations and compute the NS radius $R = R(\rho_c)$ and mass $m
= m(\rho_c)$; finally, we eliminate $\rho_c$ from these two equations,
completing our mapping of the EOS $p(\rho)$ into the mass--radius
relation $m(R)$.

Lindblom\cite{mapping} has shown how to invert the OV mapping using even a few $m(R)$ data points. Suppose that $p(\rho)$ is known up to a certain density $\rho_{max}$ from other observations and experiments.  Now start with the least dense NS, and integrate the OV equations backward, from the surface of the star [where we know $R$ and $m(R)$] down to the radius where $\rho = \rho_{max}$.  We are then left with a stellar core of known mass and radius, and we can use an analytic approximation for the solution of the OV equations to get $p_c$ and $\rho_c$; we add the point $p_c(\rho_c)$ to the EOS, and repeat for the next NS. The result is $p(\rho)$ as a piecewise linear law. This analysis can be generalized to include rotationally deformed models, and to account for the statistical uncertainty in the $m(R)$ data.  Recently, Harada\cite{Harada} has shown how other macroscopic parameters of NSs (such as moments of inertia, baryonic masses, binding energies, gravitational redshifts) can be used in this same framework to recover information about the EOS.

Finally, Saijo and Nakamura\cite{Saijo} have suggested that it might
be possible to measure the NS radius directly from the spectrum of the
GWs emitted in NS--BH coalescences.  These authors have used BH
perturbation theory to compute the spectrum of GW emitted by a disk of
dust inspiraling into a rotating BH.  When the radius
$R_\mathrm{disk}$ is larger than the wavelength of the quasi-normal
modes of the BH, the spectrum acquires several peaks with separation
$\propto R_\mathrm{disk}^{-1}$, irrespective of $M$ and $a$.  Saijo
and Nakamura conjecture that the same structure would be visible in
the spectrum of GW signals from NS--BH binaries, providing direct
information about the radius. However, several issues are left
unaddressed: in particular, the particles of the disk move along
geodesics, while the fluid of a NS would be strongly constrained by
the gravitation and pressure of the star (except, perhaps, in the
regime of severe tidal disruption); furthermore, for coalescence events
that happen at realistic distances the signal strength might be too low
to let us resolve the form-factor structure in the spectrum.

\subsection{Numerical simulations of neutron-star tidal disruption}

Summarizing, the preliminary analysis carried out by the present
author suggests that NS tidal-disruption events have much to teach us
about the NS EOS. 
This prediction is confirmed by Newtonian
simulations\cite{Centrella} of NS--BH systems, carried out using both smooth particle hydrodynamics\cite{Lee,Lee2} and Eulerian techniques\cite{Janka,Janka2}, which show that the ultimate fate of the system (complete or incomplete disruption; presence of accretion rings or tidal tails; features of the GW shutoff) depends strongly on the stiffness of the EOS.
However, detailed relativistic numerical simulations
are still needed to confirm these prospects, and will be essential as a
foundation to interpret any tidal-disruption waveforms that might be
measured in reality.

\section{Gravity waves from the $r$-modes of young neutron stars}
\label{sec:rmode}

All rotating stars possess a class of circulation modes ($r$-modes)
that are driven toward instability by gravitational radiation
reaction; in hot, rapidly rotating young NSs, this
destabilizing effect might be so strong that it dominates viscous
dissipation.  Once an $r$-mode achieves sufficient amplitude, the star
is quickly spun down as angular momentum is lost to gravitational
radiation.\cite{Lindblom,Kokkotas} For stars with initial angular velocity $\Omega \sim 1000$ Hz, the timescale for $r$-mode growth is $\sim 40$ s. In recent years, $r$-modes have attracted considerable
interest as a possible explanation for the failure to observe very
rapidly spinning pulsars, and as a promising GW source for
Earth-based detectors.

However, the astrophysical relevance of $r$-modes is still in doubt, pending judgment on two separate issues.
First, the instability of $r$-modes (discovered by analyzing the linearized Euler equations for perfect fluids) might not be confirmed after all the complicated physics that occurs in NSs is taken into account, including relativistic effects, physical-EOS effects, solid crust effects, magnetic fields, rotation laws, and so on.
Second, the amount of angular momentum removed from the star and the strength of the GW radiation emitted depend critically on the maximum amplitude that can be reached by the $r$-mode;
but the growth of the $r$-mode might be limited to a very small \emph{saturation amplitude} by the effects of magnetic fields\cite{Rezzolla}, or by the leakage of energy to other (damped) modes by way of nonlinear hydrodynamical couplings.

There have been several attempts to investigate the nonlinear dynamics of the $r$-modes, by means of second-order Lagrangian perturbation theory,\cite{Teukolsky} and of relativistic numerical simulations in the Cowling approximation.\cite{Stergioulas} In addition, Lindblom, Tohline and the present author have carried out Newtonian numerical simulations that include radiation reaction as an effective force. We now briefly review this work, but we invite the reader to refer to the original articles\cite{rmode1,rmode2} for further details.

\subsection{Numerical evolutions of nonlinear $r$-modes in neutron stars}

We solved the Newtonian Euler and Poisson equations on a $128 \times 64
\times 128$ cylindrical grid, using a 2nd-order--accurate
finite-difference code developed at LSU\cite{LSU} to tackle a
variety of astrophysical problems. The LSU code was parallelized\cite{Motl} in the domain-decomposition paradigm using the well known MPI\cite{MPI} library.  We ran it on 16 nodes of the HP V-2500 supercomputers at Caltech's Center for Advanced Computing Research.\cite{CACR}

Our initial equilibrium configuration was a simple polytrope (with
$n=1$, $M=1.4 \, M_\odot$, $R = 12.5$ km) obtained by solving
selfconsistently the Bernoulli and Poisson equations.\cite{Hachisu}
We then augmented the rigid-rotation velocity field with a small-amplitude, slow-rotation approximation to an $l = m = 2$ $r$-mode,\cite{Lindblom98}
\begin{equation}
\delta \vec{v} = \alpha_0 R \Omega_0 \Bigl( \frac{r}{R} \Bigr)^2
\vec{Y}^{B}_{22} e^{i \omega_0 t},
\label{eq:approxmode}
\end{equation}
where $R$ and $\Omega_0$ are the radius and angular velocity of the
unperturbed star, $\alpha_0$ is the dimensionless $r$-mode
amplitude, and $\vec{Y}^{B}_{22}$ is a vector spherical harmonic
of the magnetic type.  Gravitational radiation reaction was added as an effective Newtonian force\cite{force} proportional to the derivatives of the current multipole moments of the star.

Our computing budget set tight limits on the length of the simulation: one full rotation period of a rapidly rotating NS took up to ten thousand CPU hours. Consequently, we could not afford to follow the growth of the $r$-mode through one or more radiation timescales.
Instead, we increased the strength of the effective radiation-reaction force by a factor of about 4500, to bring down the $r$-mode growth timescale to values comparable to the rotation period.
Even with this trick, we expect that the physical behavior observed in the simulations should still be realistic, because the nonlinear dynamics of the $r$-mode (including its couplings to other modes) should happen at the hydrodynamical timescale, which in our simulations is still much shorter than the radiation-reaction timescale.
\begin{figure}
\begin{center}
\includegraphics[width=3.3in]{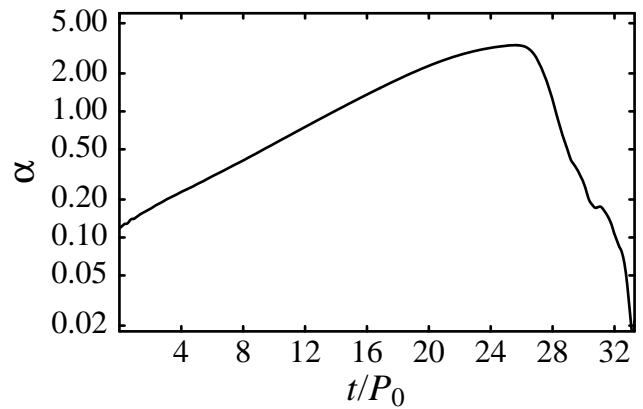}
\caption{Evolution of the dimensionless $r$-mode amplitude $\alpha$. Amplitudes of order one imply mode velocities comparable to the local velocity of rigid rotation. Time is measured in units of the initial rotation period $P_0$ of the NS.\label{fig:alpha}}
\end{center}
\end{figure}

While computing the evolution of the star, we monitored the $r$-mode amplitude $\alpha$ by projecting the velocity of the fluid on $\vec{Y}^B_{lm}$,
\begin{equation}
\alpha = \frac{2 |J_{22}|}{(\Omega/R) \int \rho r^6 dr},
\end{equation}
where $J_{22}$ is the $l = m = 2$ current multipole moment;
similarly, we read off the mode frequency $\omega$ as
\begin{equation}
\omega = - \frac{1}{|J_{22}|}\left|\frac{d{J}_{22}}{dt}\right|.    
\end{equation}
At the beginning of our evolution, these diagnostics recovered the theoretical values $\alpha_0$ and $\omega_0$ within the expected error.

Figure \ref{fig:alpha} shows the evolution of the $r$-mode amplitude. For this simulation, the theoretical prediction\cite{Lindblom98} for the $r$-mode growth time was 10 initial rotation periods. The mode grew exponentially (in good accord with theory) until $\alpha \sim 2$; then the mode started to be limited by some nonlinear process, the amplitude peaked at $\alpha = 3.35$, and finally fell down very rapidly.  A movie of the crashing $r$-mode can be found on the web.\cite{vallis}
\begin{figure}
\begin{center}
\includegraphics[width=3.3in]{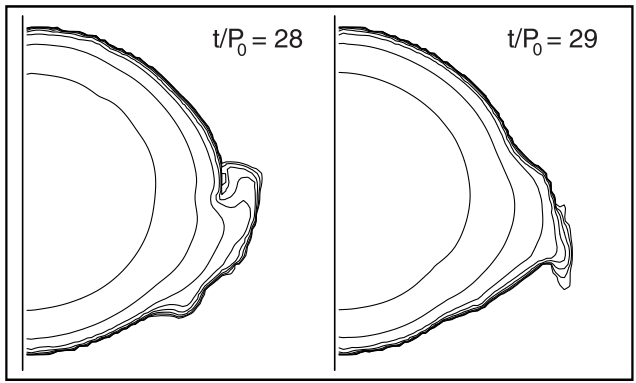}
\caption{Cresting waves at the surface of the NS, after the $r$-mode amplitude has reached its maximum. The waves create strong shocks that dump mode energy into thermal energy.\label{fig:shock}}
\end{center}
\end{figure}

What nonlinear process was responsible for limiting the growth of the $r$-mode, and for causing its rapid demise?
We discovered a clue when we examined the evolution of the energy and angular momentum in the course of the simulation.  Even after the emission of angular momentum into GWs fell to zero, the star continued to lose energy; something other than GWs must be responsible for this loss.

We believe that we have found the culprit.
To first order in the amplitude, the $r$-mode is only a \emph{velocity} mode; to second order, however, there is an associated density perturbation, proportional to $Y_{32}$, which appears as a wave with four crests (two in each hemisphere) on the surface of the star. As the amplitude reaches its maximum, these propagating crests turn into large, breaking waves (see Fig.\ \ref{fig:shock}), and the edges of the waves develop strong shocks that dump kinetic energy into thermal energy, killing the $r$-mode.

\subsection{Numerical simulations as a laboratory of stellar physics}

Our code provided a nice laboratory to perform several more
evolutions and tests.
\begin{enumerate}
\item We performed basic tests on the robustness of the code and of our diagnostics, evolving stars with different angular velocities, with or without $r$-mode perturbations, and at different grid resolutions.
\item We investigated the dependence of the saturation
amplitude on the artificial amplification of radiation reaction. Our computing budget made it impossible to increase the radiation-reaction timescale; instead, we reduced it even more, finding that the $r$-mode saturates faster, but at essentially the same amplitude.
\item We studied the \emph{unforced} evolution of unit-amplitude $r$-modes, finding that they are essentially stable for as long as we could evolve them. These results are compatible with the relativistic evolutions (also unforced) performed by Stergioulas and Font.\cite{Stergioulas}
\end{enumerate}

\subsection{A new picture for gravity-wave signals from $r$-modes}

In the traditional scenario for GW emission from $r$-modes,\cite{Owen98} the unstable $r$-mode would grow until it reached a dimensionless amplitude of about one, and then it would saturate and persist at that amplitude (for several months) until it would have lost most of its angular momentum; during that time, the frequency of the $r$-mode would decrease in proportion with the NS spin.
The prospective picture of $r$-mode GW signals that emerges from our evolutions is quite different.  Most interesting, the $r$-mode spindown episodes are faster (only a few minutes), and the GW frequency remains remarkably constant as the angular velocity of the star decreases.  As a result, the search for $r$-mode signals changes from a pulsar-like search (which must account for the Doppler shifts generated by the movement of the Earth around the sun) to an easier chirp-like search, with encouraging prospects for detection.\cite{Lindblom01d}

Although recent studies of the effect of magnetic fields,\cite{Rezzolla} and of exotic forms of bulk viscosity\cite{Lindblom01e} suggest
that the astrophysical significance of $r$-modes might be limited, the considerable uncertainty about the macroscopic and microscopic structure and dynamics of NSs makes it reasonable to devote some resources to GW searches for $r$-mode signals similar to those predicted by the simulations described above.

\begin{acknowledgments}
For support, inspiration, and very useful discussions, the author thanks Kip Thorne, Lee Lindblom, Joel Tohline, Massimo Pauri and Luca Lusanna. This research was supported by NSF grants PHY-0099568 and PHY-9796079.
\end{acknowledgments}

\end{document}